\documentclass[%
 aip,
 amsmath,amssymb,
 reprint,%
]{revtex4-1}

\usepackage{graphicx}
\usepackage{dcolumn}
\usepackage{bm}

\usepackage[utf8]{inputenc}
\usepackage[T1]{fontenc}
\usepackage{mathptmx}
\usepackage{etoolbox}

\makeatletter
\def\@email#1#2{%
 \endgroup
 \patchcmd{\titleblock@produce}
  {\frontmatter@RRAPformat}
  {\frontmatter@RRAPformat{\produce@RRAP{*#1\href{mailto:#2}{#2}}}\frontmatter@RRAPformat}
  {}{}
}%
\makeatother
\begin{document}

\preprint{AIP/123-QED}

\title{Phase behaviour and dynamical features of a two-dimensional binary mixture of active/passive spherical particles}

\author{Diego Rogel Rodriguez}
\affiliation{Departamento de Estructura de la Materia, Fisica Termica y Electronica, Universidad Complutense de Madrid, 28040 Madrid, Spain}
\affiliation{GISC - Grupo Interdisciplinar de Sistemas Complejos 28040 Madrid, Spain. }

\author{Francisco Alarcon}
\affiliation{Departamento de Estructura de la Materia, Fisica Termica y Electronica, Universidad Complutense de Madrid, 28040 Madrid, Spain}
\affiliation{GISC - Grupo Interdisciplinar de Sistemas Complejos 28040 Madrid, Spain. }

\author{Raul Martinez}
\affiliation{Departamento de Estructura de la Materia, Fisica Termica y Electronica, Universidad Complutense de Madrid, 28040 Madrid, Spain}
\affiliation{GISC - Grupo Interdisciplinar de Sistemas Complejos 28040 Madrid, Spain. }
\affiliation{Departamento de F\'isica Te\'orica de la Materia Condensada, Facultad de Ciencias, Universidad Aut\'onoma de Madrid, 28049 Madrid, Spain. }

\author{ Jorge Ram{\'i}rez}
\affiliation{Departamento de Ingenier{\'i}a Qu{\'i}mica, ETSI Industriales, Universidad Polit{\'e}cnica de Madrid, 28006 Madrid, Spain. }

\author{  Chantal Valeriani}
\affiliation{Departamento de Estructura de la Materia, Fisica Termica y Electronica, Universidad Complutense de Madrid, 28040 Madrid, Spain}
\affiliation{GISC - Grupo Interdisciplinar de Sistemas Complejos 28040 Madrid, Spain. }


\begin{abstract}
In this work we  have characterized the phase behaviour and the dynamics of bidimensional mixtures of active and passive Brownian particles. 
  We  have evaluated state diagrams at several concentrations of the passive components 
  finding that, while passive agents tend to hinder phase separation,  active agents force crystal-like 
  structures on passive colloids. 
In order to study how passive particles affect the dynamics of the mixture, we have computed the long-time 
diffusion coefficient of each species, concluding that  active particles induce activity and super-diffusive
   behaviour on passive ones.
Interestingly, at the density at which the system enters a MIPS state the active particles' diffusivity 
shows an inflection point and the passive particles' one goes through a maximum, due to the change in the dynamics of the active components, as shown in the  displacement's probability distribution function. 

\end{abstract}






\maketitle

\subsection{Introduction}
\label{sec:intro}
Active matter systems are composed of constituents that consume energy in order to move or exert mechanical forces, 
constantly driving themselves away from equilibrium \cite{IntroRef}.
Examples of active particles at the mesoscopic scale are living,
such as bacteria, or artificial, such as active  colloids  \cite{exp1, exp2}.
 In these systems, the violation of time reversal symmetry, 
which is the main feature of non-equilibrium dynamics, occurs at the level of individual components, 
unlike driven systems such as granular matter or thermophoretic colloids, 
in which the departure from equilibrium is due to external fields or 
boundary conditions. 
One of the characteristic features of active matter is collective motion, i.e. a behaviour in 
which the action of an individual unit is dominated by the influence of other components. 
The collective behavior  can be originated from the deviation from equilibrium, leading to novel phenomena such as nonequilibrium phase transitions or large scale directed motion \cite{collective}.



Experiments on spherical man-made self-propelling colloids 
have shown that active particles can present interesting emergent collective properties \cite{exp3, exp4,mips}, such as motility-induced phase separation (MIPS). This phase separation involves the spontaneous assembly of particles due to the persistence of their direction of  motion, forming a stable dense and dilute phase even in the absence of an explicit attraction   or aligning interactions \cite{comp1}. 
An example of active particles   undergoing MIPS under suitable conditions are Active Brownian Particles (ABP), i.e. self-propelled Brownian particles that could interact with each other via a purely repulsive potential.
To understand the basic microscopical mechanism underlying MIPS in ABP, 
it is important to study the time it takes a particle to reorient its direction of motion with respect to the thermal noise\cite{bechinger,reviewmarchetti,reviewcacciuto}. 

Systems formed by a mixture of active and passive Brownian colloids  provide a novel way 
for switchable self-assembly \cite{switch}, micro-rheological measurements \cite{rheology}, or help understanding active dynamical processes within the cell \cite{cell}. 
For this reason, their structural properties and phase behaviour have  been recently the subject of a few studies, mostly focused on the formation of MIPS \cite{mix1, mix2}.


The goal  of our work is to characterise both the phase behaviour and the dynamics of  a two dimensional binary mixture of active and passive Brownian particles. 
For the former, we will numerically study a mixture of monodisperse active/passive spheres,  
to establish phase separation  within a broad parameter space
(density, P\'eclet and fraction of passive particles).  
For the latter, we will compute the diffusion coefficient of both passive and active colloids
together with the displacement's probability distribution function.  
Our findings can be considered  relevant to the design of experimental devices that pursue the driven self-assembly of passive colloids, such as the ones used in microrheological measurements \cite{rheology}, or shed light upon the behaviour of active matter in heterogeneous media. 

The manuscript is organised as follows. In section II
we present the simulation details and 
the tools used to characterise both structure and dynamics. 
In section III we present our results for the structural properties and in section IV for the dynamical features. 
Finally, we discuss our conclusions in section V.

\section{Numerical Methods}

In order to study the  binary mixture of  active/passive particles, we perform  
Brownian Dynamics  simulations with an in house modified version of the {\it LAMMPS}\cite{LAMMPS}  open source package.
The system consists of  
  $N=N_a+N_p$ spherical colloids in a two-dimensional box of size $L$ with periodic boundary conditions,  
where $N_a$ is the number of active colloids and $N_p$ is the number of passive colloids. It is a monodisperse mixtures, with diameter $\sigma$. 

All particles undergo a translational Brownian motion (eq.\ref{eq:motion}),  where active particles are 
characterised by an additional constant self-propelling force $F_a$ acting along the orientation vector $\vec{n}_i$.
Besides translation, the orientation vector of active particles evolves according to a rotational diffusion equation (eq. 2).
\begin{align}
\label{eq:motion}
& \dot{\vec{r}}_i = \frac{D}{k_B T} \left( - \sum_{j\neq i} \nabla V(r_{ij}) + | F_a |\, \vec{n}_i \right) + \sqrt{2D} \, \vec{\xi}_i, \\
& \dot{\theta}_i = \sqrt{2D_r}\, \xi_{i,\theta},
\end{align}
where $V(r_{ij})$ is the interparticle pair potential, $k_B$ is the Boltzmann constant, $T$ is the temperature, and the components of $\vec{\xi}_i$ and $\xi_{i,\theta}$ are white noise with zero mean and correlations $\langle\xi^{\alpha}_i(t)\xi^{\beta}_j(t')\rangle = \delta_{ij} \delta_{\alpha\beta} \delta(t-t')$. 
The translational, $D$, and rotational, $D_r$, diffusion coefficients obey the Stokes-Einstein relation \cite{StokesEinstein}: $D_r = 3D/\sigma^2$.  
Both passive and active particles interact with each other via a purely repulsive WCA potential \cite{WCA}.
\begin{eqnarray}
V(r)  = \begin{cases} & 4 \epsilon \left( \left( \dfrac{\sigma}{r} \right)^{12} -  \left(\dfrac{\sigma}{r} \right)^6 \right) + \epsilon, \quad  r < 2^{1/6} \sigma \\ 
& 0, \qquad \qquad \qquad \qquad \qquad \quad   \; r \geq 2^{1/6} \sigma
\end{cases}
 \label{eq:WCA}
\end{eqnarray}
where $\epsilon$ has energy dimensions, $r$ is the center to center distance, and $\sigma$  the particles' diameter.
 Throughout this study, we have used  reduced units, in which lengths, times and energies are given in terms of $\sigma = 1$, $\tau = \sigma^2/D$,  
  $k_B T = 1$ and $D=1$, and set {\bf $\epsilon = k_B T$}.
  The time step  has been set to $\Delta t = 10^{-5} \tau$.

  As a measure of the degree of  activity, we use the P\'eclet number, an adimensional number defined as 
\begin{equation}
\text{Pe} =  \frac{3v}{\sigma D_r},
\end{equation}
where $v=|F_a|D/k_BT$ is the self-propelling velocity.
 If Pe $\ll 1$, the particle's motion is governed by diffusion, as for passive Brownian particles, while for Pe $\gg 1$ diffusive transport is negligible compared to active motion.

 \begin{figure}[h!]
\includegraphics[width=\columnwidth]{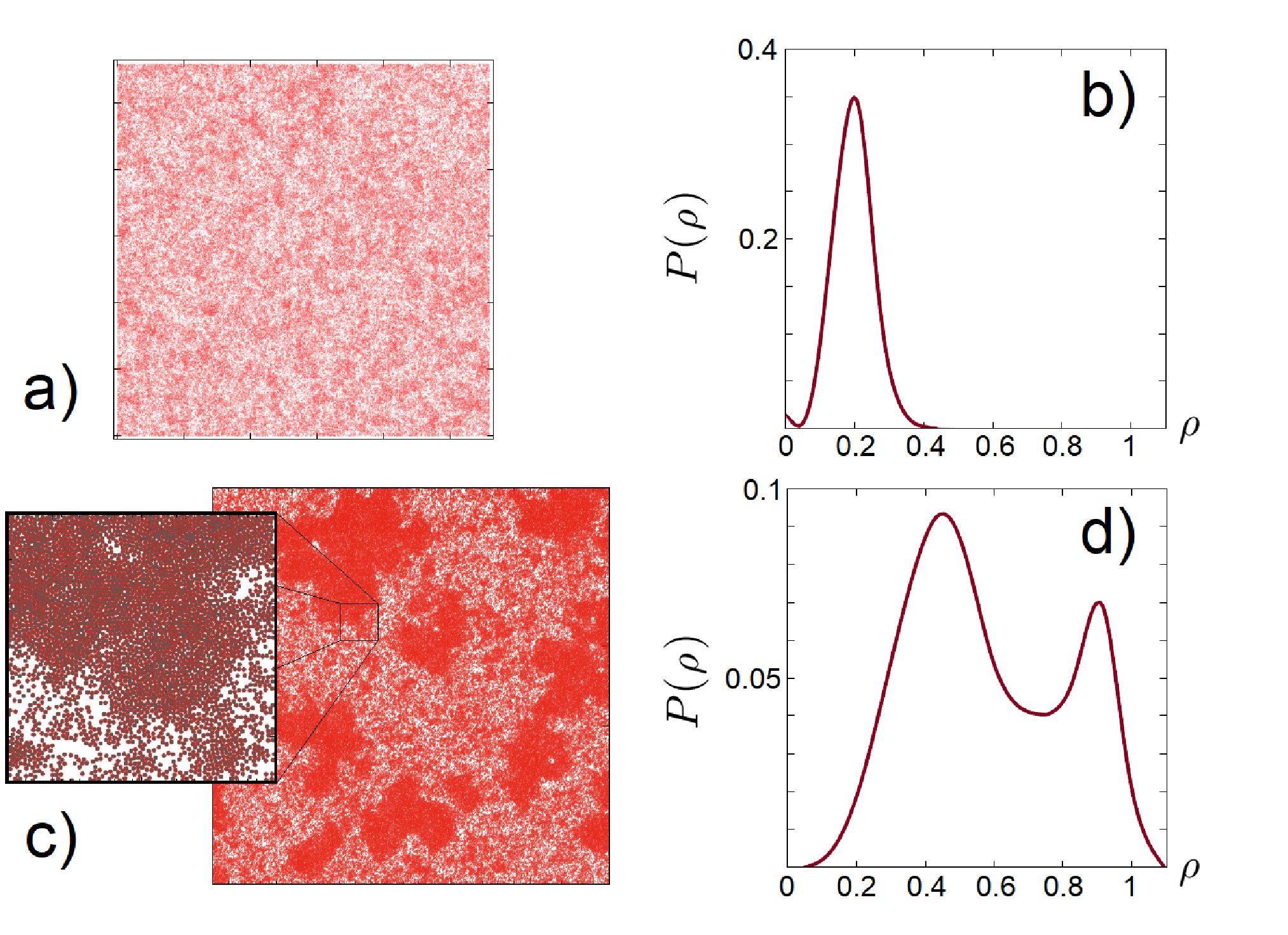} 
\caption{Snapshots of ABP at $\rho=0.5$ and $N_p=0$  showing  a) no phase separation (Pe=40), b) its $P(\rho)$ distribution $P(\rho)$ is centered at $0.2$ and  c) MIPS (Pe=120) and d) its  $P(\rho)$ distribution.}
\label{fig:pphi}
\end{figure}
In this study we systematically change the Pe number, the density of particles ($\rho=N/(L^2)$) and the ratio between active and passive particles. We choose to work with the total number density $\rho$ (instead of the area fraction $\phi$), 
as it is a magnitude that does not depend on the particle's diameter, as we will comment later in the text.

When focusing on motility induced phase separation in  a monodisperse mixture we consider system's sizes ranging from $N=90000$ to $N=180000$, setting 
the box lenght $L$ accordingly  to obtain the  desired total number density. 



 In order to detect phase separation, we have implemented a standard Voronoi cell method to measure local number density \cite{Voronoi}.
  This method involves computing a Voronoi tessellation over the simulation box, where each colloid is in a corresponding Voronoi cell and the local number density $\rho$ is given as the inverse of the area of each cell.
  We define $P(\rho)$ as 
  the probability distribution of finding a cell with local number density $\rho$. 
 Typically, an isotropic system (Fig.~\ref{fig:pphi}a)
 shows a $P(\rho)$ with only one local maximum around the total number density $\rho$ (Fig.~\ref{fig:pphi}b), 
 whereas a phase separated system (Fig.~\ref{fig:pphi}c) shows two local maxima: one centred at the dilute phase and another at the concentrated phase (Fig.~\ref{fig:pphi}d). 
 Thus, an inspection of $P(\rho)$ will reveal the existence of a motility induced phase separation. 




In order to characterize the dynamics of the system, we  have computed the long time diffusion coefficient via the 
 mean-square displacement (MSD) of either active or passive particles 
 in the presence of active ones, given by
\begin{equation}
\text{MSD}_{a/p}(t) = \frac{1}{N_{a/p}} \sum_{i=1}^{N_{a/p}}\left(\vec{r}_{a/p,i}(t)  - \vec{r}_{a/p,i}(0)  \right)^2.
\end{equation}
\noindent
Moreover, we have also computed the probability distribution function of  particles' displacements for either passive and active particles.



\section{Results}

\begin{figure}[h!]
\includegraphics[width=\columnwidth]{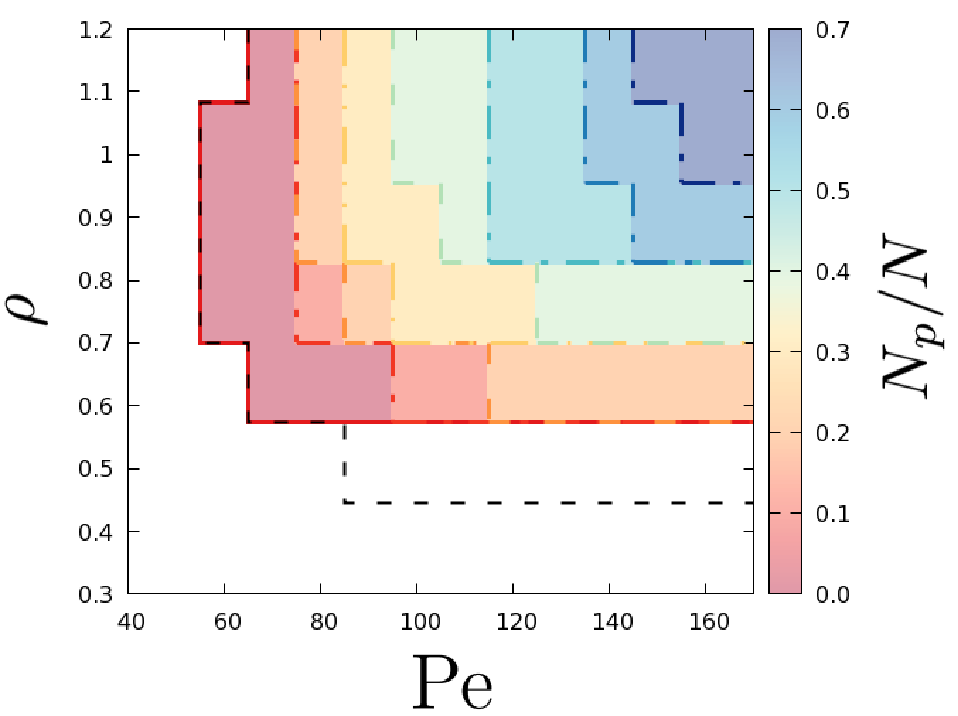} 
\caption{$\rho$-Pe state diagrams of monodisperse mixtures of active/passive particles. Coloured areas correspond to  MIPS, different colours representing different fractions $N_p/N$ of passive components.  Black dashed line is for purely active WCA particles from Ref. \cite{Stenhammar}.}
\label{fig:phase}
\end{figure}

To unravel the role played by  passive colloids in a suspension of active agents
we estimate  in Fig. \ref{fig:phase} $\rho$-Pe state diagrams for a wide range of P\'eclet numbers, 
total number densities $\rho$, and fraction of passive particles $N_p/N$.  
The reason behind using the total number density instead of the packing fraction 
has been already  suggested in Ref.\cite{Stenhammar,reviewpe2}: 
when changing the propulsion force implies changing the Peclet number, 
when increasing Pe particles  end up colliding at a very high speed, thus 
decreasing their effective diameter.


As shown in Fig.\ref{fig:phase}, we mostly focused on relatively large Peclet numbers (larger than 40). 
Interestingly, for lower values of Pe  the authors  of  Ref.\cite{newref} demonstrated the existence of a coexistence 
between an active liquid and a hexatic phase in  a two dimensional suspension 
of  Active Brownian Particles. . 

For purely active systems (dashed black line) and  systems with the lowest fractions of passive particles (in red), 
MIPS are formed when densities are at least  $\rho \sim 0.5$ 
    (corresponding to $\phi \sim 0.4$) and Pe ranging from 60 (at the highest densities) 
    to 80 (at the lowest densities). 
    When comparing to the state diagram presented by Stenhammar \textit{et al.} \cite{Stenhammar}
    (black dashed line in the figure), we detect a small difference 
      when $\rho \sim 0.5$ 
      ($\phi \sim 0.4$)
This is due to the different definition of  Peclet number adopted in our work and in their work:   while in our work 
$D_r$
is set to 3
 and the self-propelling 
   velocity is changed in order to vary the P\'eclet number, Stenhammar \textit{et al.}.
   vary Pe by fixing $v$ and varying $k_BT$ and so $D_r$. 
   As we will present  below, at high P\'eclet numbers, our definition effectively reduces the 
   particle diameter and, therefore, the area fraction of particles in the system.
   However, when following the same procedure as Stenhammar, 
   we recovered their results, 
   with MIPS at
    number densities of $\rho \sim 0.5$ and P\'eclet =90 and above (see Appendix).

Adding passive components shifts the state diagram to higher values of Pe and $\rho$ simultaneously:  
  higher densities and  activities are needed to get phase separated systems. 
  Even though passive components tend to hinder phase separation, MIPS 
 are detected  even in systems containing up to $70 \%$  of passive particles, whenever the self-propelling velocity of 
 active components and the total density are high enough. 
 On the contrary, MIPS have not been observed at higher fractions of passive agents in the
  range of P\'eclet numbers considered in this work (up to $\text{Pe} = 250$).

The authors  of Ref. \cite{mix1}   studied 
mixtures of active and passive particles and 
derived a relationship between  density and  Pe number ($\phi_0 \sim 1/Pe$)  using the kinetic model developed in \cite{analytical}: they concluded that this relationship 
was capable of  fitting the results  for ABP simulations when  $\phi>0.4$ and  $N_p/N = 0.5$.
We observe that all MIPS regions can be scaled one on top of another. 
  \begin{figure}[h!]
\includegraphics[width=\columnwidth]{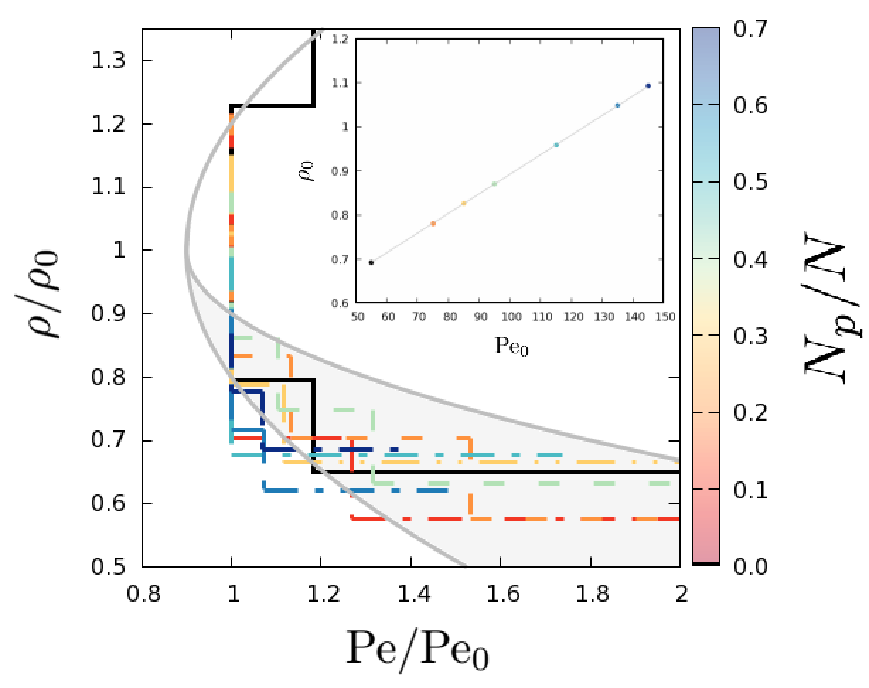} 
\caption{State diagrams  (coloured lines) as in Fig. \ref{fig:phase} where $\rho$ and Pe have been 
normalized by the estimated  values of $\rho_0$ and Pe$_0$ for each $N_p/N$.  
Black line (purely active case). 
Gray curves: dispersion
estimated fitting  the state diagrams to a binodal curve. Inset: linear behaviour of  $\rho_0$ versus Pe$_0$.}
\label{fig:critical}
\end{figure}

Differently from  Ref. \cite{mix1}, the MIPS boundaries in Fig. \ref{fig:phase}  have a similar parabolic-like shape 
 culminating in what could be identified as a critical point (Pe$_0$, $\rho_0$) 
for every value of $N_p/N$. 
 With this idea in mind, we approximately estimate the location of 
 Pe$_0$ and $\rho_0$ for every fraction $N_p/N$ of 
passive particles. 
    Fig. \ref{fig:critical} shows the normalized phase diagrams
of monodisperse mixtures when the curve for the purely active case (black line, $N_p/N=0$)
is taken as reference.  All curves seem to collapse 
onto a master curve when representing them as a function of  $\rho/\rho_0$ vs Pe/Pe$_0$.


Interestingly, the critical points $\rho_0$, Pe$_0$ for the different passive
  fractions are aligned,  as shown in the inset of Fig. \ref{fig:phase}. 
  On the one side, this might allow us to predict, for every concentration of passive particles up to close packing, 
  the maximum Peclet beyond which MIPS can be detected (being $\rho=1.15$ the maximum close packing density). 
On the other side, this suggests that the normalised state diagrams for  
active/passive binary mixtures are very similar at every  $N_p/N$,  for mixtures with up to 70\% of passive particles. 
Beyond that ratio, the overall density needed to observe MIPS is so high that the mixture 
behaves as a very dense suspension of passive particles,
where particles interact repulsively and  MIPS are absent.
Therefore, given that the rescaled state diagrams for low/moderate fractions of passive particles are similar,
one might wonder whether the structure of MIPS (whenever it occurs) may also be similar.

 \begin{figure}[h!]
\includegraphics[width=0.8\columnwidth]{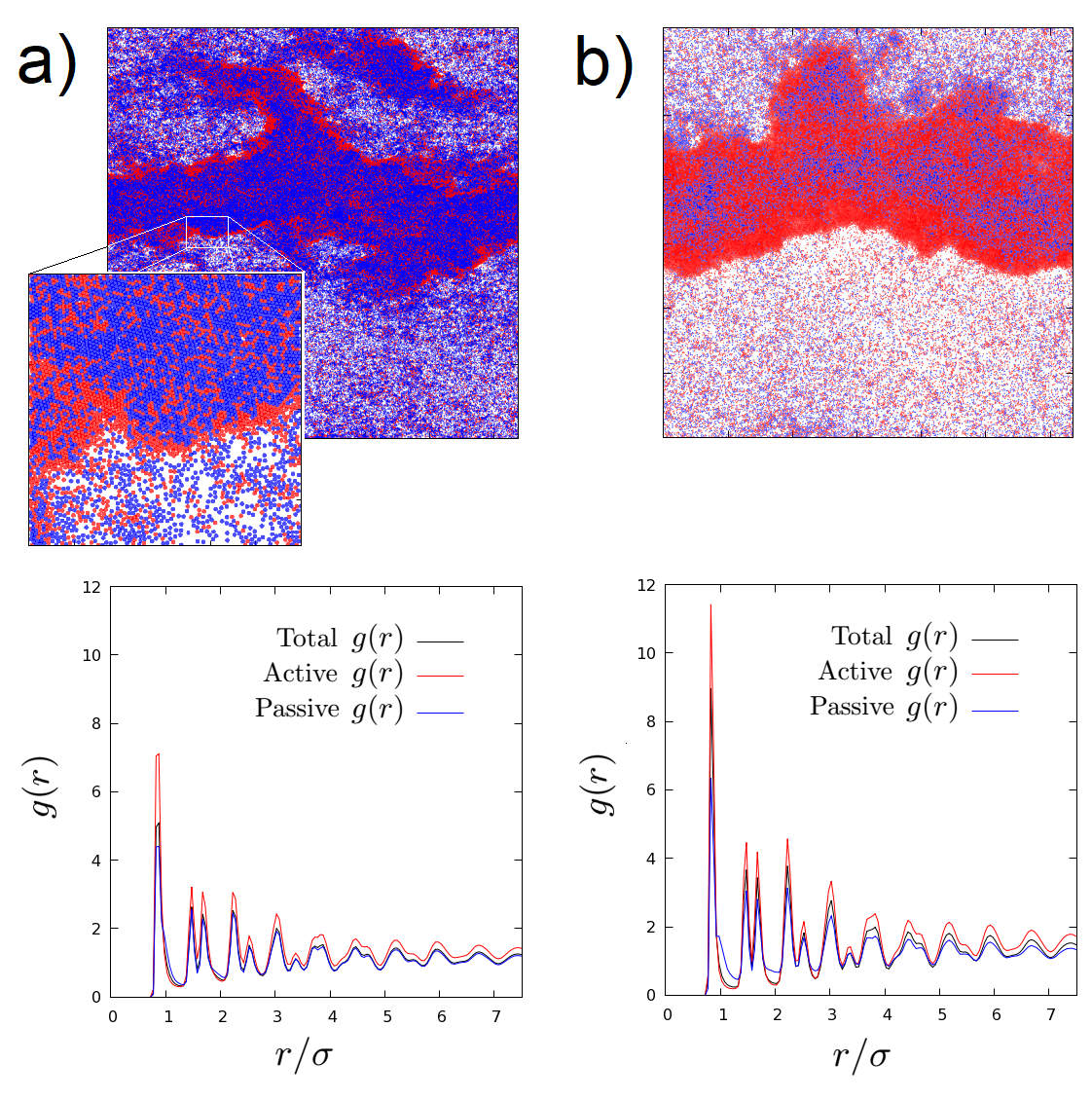} 
\caption{ Top: snapshots  of monodisperse phase-separated
 mixtures of passive (blue)/active (red) colloids:  a) $N_p/N = 0.6$, $\rho = 1.018$, $\text{Pe}=190$, b) $N_p/N = 0.3$, $\rho = 0.891$, $\text{Pe}=170$. 
Bottom:  radial distribution functions corresponding to the top panels.}
\label{fig:clustering}
\end{figure} 

   \begin{figure*}
  \centering \includegraphics[width=\textwidth]{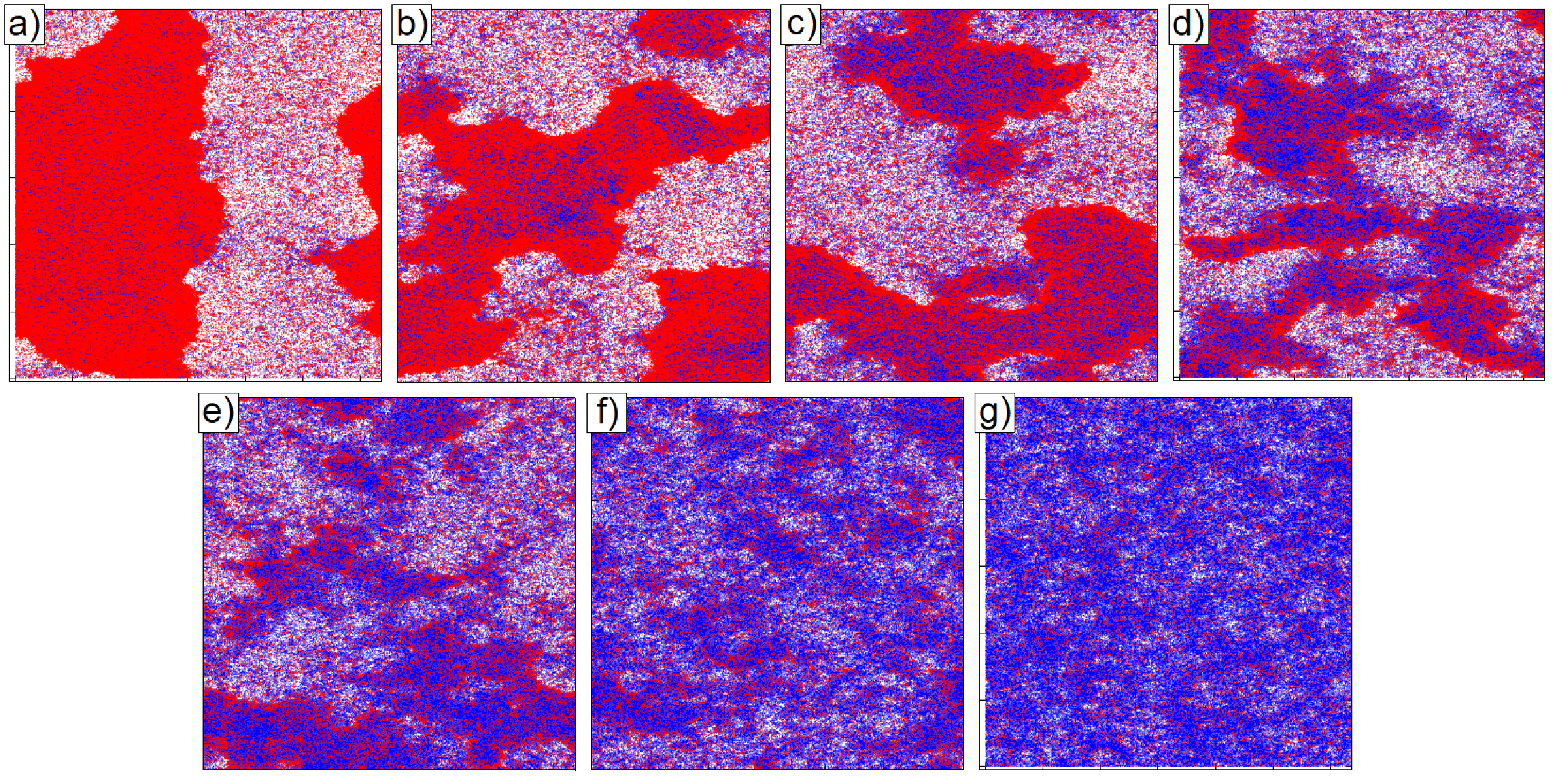}
  \caption{ a-g) Snapshots  at $\rho = 1.018$, Pe $=120$ and several $N_p/N$ fractions (passive particles in blue and    active ones in red): a) $N_p/N=0.1$ b) $N_p/N=0.2$ c) $N_p/N=0.3$ d)$N_p/N=0.4$ e) $N_p/N=0.5$ f) $N_p/N=0.6$ g) $N_p/N=0.7$.  a), b), c), d), e) and f) show MIPS, while in g), the high $N_p/N$  forbids MIPS. Movies of panel b), d) and e) can be found in the Supplementary Information.}
  \label{fig:morfologias}
\end{figure*} 

In the top panels of Fig. \ref{fig:clustering}, we compare the snapshots of two different MIPS cases, 
corresponding to a) high passive particle concentration ($N_p/N = 0.6$, $\text{Pe}=190$, $\rho = 1.018$, 
deep into the blue region of Fig. \ref{fig:phase}) and b) low passive particle concentration 
($N_p/N = 0.3$, $\text{Pe}=170$, $\rho = 0.891$, deep into the yellow region of Fig. \ref{fig:phase}).
 In either case,  the denser phase is formed by a skin of active colloids (in red) encapsulating the passive ones (in blue),  
 even when the  passive particles' concentration is lower  (panel b). 
When analysing the phase-separated mixture in more detail, 
we observe that the  self-assembly of the dense phase
 is triggered by active fronts that trap the passive components, 
 favouring the spontaneous aggregation of passive agents. 
 However, in the dilute phase  the 
 distribution of passive and active particles is more homogenous.


As shown by the 
 partial pair distribution function $g(r)$ for active and passive colloids (Fig. \ref{fig:clustering}-bottom panels), 
 crystal-like structures are present at short 
 distances, and liquid-like  at longer distances.
 However, the partial $g(r)$   for active particles reveals a structure that is better 
 defined than the one  for passive agents. Given that 
 passive  particles would never show the same structuring in the absence of 
 active agents, one could conclude that 
active particles 
 force   clustering of passive ones, as if an effective attraction was present between them.
 The partial radial distribution functions for  active/passive colloids show a first high peak, indicating 
 that both types of colloids are strongly packed, at $ r < \sigma$.
The reason is that  in order to alter the Pe we change  the self-propelling speed of the active particles.
This reduces  the effective particle radius when  activity increases, 
being the WCA a not-too stiff interaction potential.  
 This result motivated our choice of adopting $\rho$ as the measure of density, instead of the area fraction used in 
Ref.\cite{Stenhammar}.

 
 In order to better unravel possible differences between MIPS at different concentration of passive particles, 
  we  focus on one  point of the state diagram (Pe $=120$, $\rho = 1.018$ in Fig. \ref{fig:phase}) 
 where most of the different  $N_p/N$ fractions show MIPS.  
 Fig. \ref{fig:morfologias} shows a set of snapshots 
for several fractions of passive particles, ranging from the lowest $N_p/N$ (panel a) to 
the highest $N_p/N$ (panel g). While in most  cases  the state point is well
 in the MIPS region ($N_p/N=0.1$ (panel a),    $N_p/N=0.2$ (panel b),   $N_p/N=0.3$ (panel c),  
 $N_p/N=0.4$  (panel d),  $N_p/N=0.5$ (panel e), $N_p/N=0.6$ (panel f)), the case corresponding to 
 $N_p/N=0.7$ (panel g) lies   outside  its corresponding MIPS boundaries. 
  We note that, when increasing the amount of passive particles, 
  the system tends to form many smaller clusters
until a maximum threshold for $N_p/N$ is reached, for which the active agents can no longer trap passive particles
and are unable to trigger MIPS.
 Furthermore,   density fluctuations in these mixtures are larger  than in the purely active systems. 
 While purely active systems  maintain 
 a relatively stable phase separation (as shown in panel a,  very close to a purely active case), clusters in mixtures of active and passive colloids 
 are more dynamic, since they are permanently moving, vanishing and remerging (see  videos in the supplementary information).

   When studying the local number density distribution functions $P(\rho)$ of  active/passive 
particles associated with MIPS, we notice an interesting behaviour in the way 
the $P(\rho)$  evolves as a function of $\rho$ and Pe. 
\begin{figure}[h!]
\includegraphics[width=1.05\columnwidth]{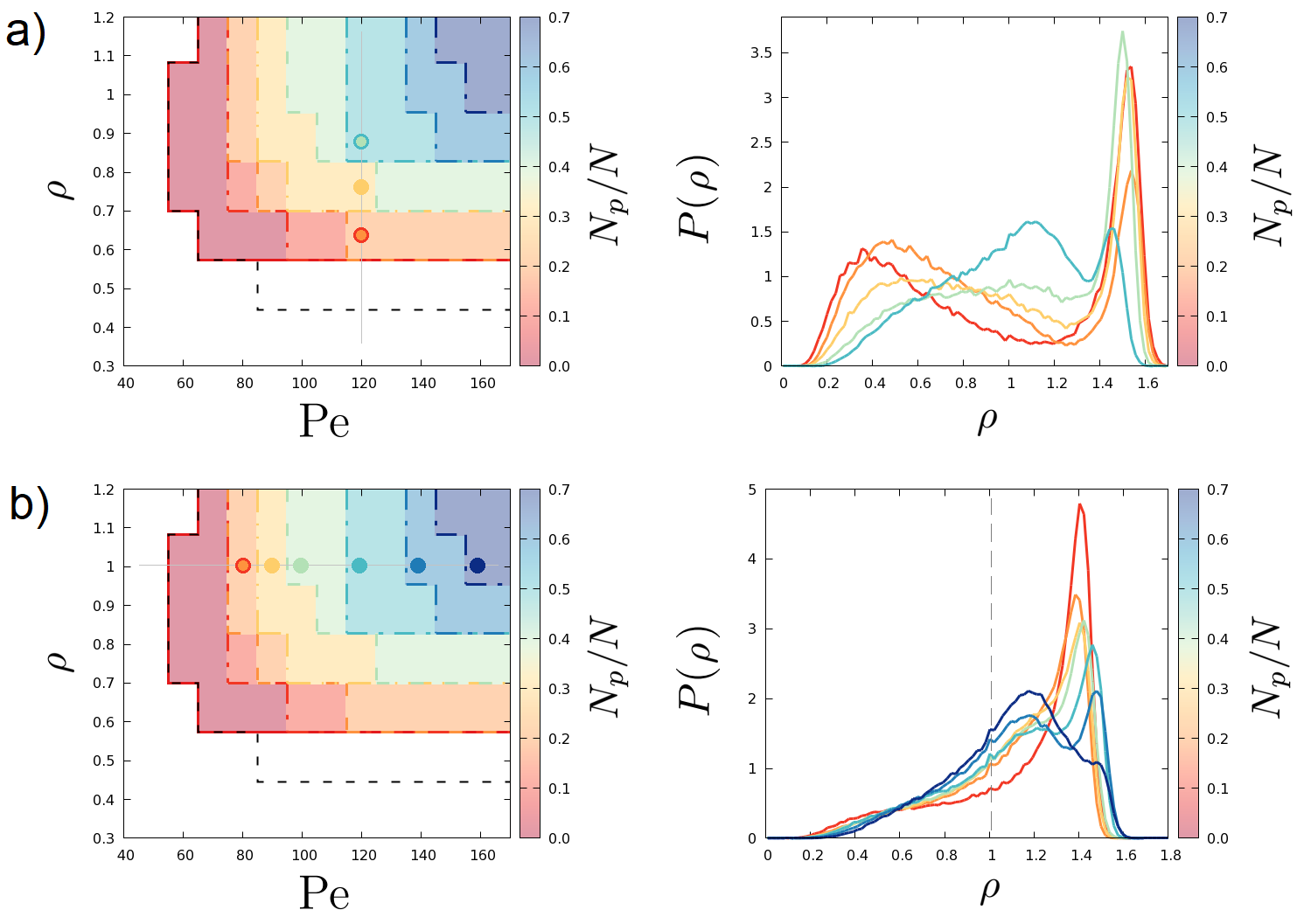} 
\caption{Right panel: local number density distributions $P(\rho)$ for  systems at the MIPS boundaries  for a) a constant P\'eclet number of Pe$=120$, b) a constant number density of $\rho=1.018$. Left panel: phase diagrams of monodisperse mixtures of active/passive colloids, with coloured circles indicating the $\rho$ and Pe conditions  for which its corresponding $P(\rho)$ is shown.}
\label{fig:mixphase}
\end{figure}
While  a system consisting of only passive particles 
 (without active agents)
    presents a  gaussian $P(\rho)$ distribution  centred around the 
 total number density, the shape of the $P(\rho)$ can be strongly affected by the amount of active particles added
 to the suspension.
 Fig. \ref{fig:mixphase}(a) shows a 
series of $P(\rho)$  (right panel) of  systems at different nominal densities, each one next to the MIPS boundaries,  depending on  
$N_p/N$  and at a fixed Pe of $120$ (left panel).  
  For purely active (red) and systems with $N_p/N$ up to about 0.5  (light green) a clear phase separation appears, 
 with two well defined peaks at low and high densities, respectively.  
 This corresponds  to the encapsulating mechanism
  in  MIPS  (observed in Fig.\ref{fig:clustering} and  \ref{fig:morfologias}), in which  passive particles are trapped in a
   dense phase by the active colloids, while the rest of them remain in a  low density phase. 
 Increasing $N_p/N$ even further leads to a less pronounced phase separation, 
 with the two peaks merging each other around the nominal total density (as shown in panels e and f of 
 Fig. \ref{fig:morfologias}).

   Fig. \ref{fig:mixphase}(b) shows a series of $P(\rho)$  (right panel)  for phase separated systems, 
   fixing $\rho = 1.018$  and varying Pe (left panel).  We notice that, as $N_p/N$ increases 
   the distance between the two local maxima in $P(\rho)$ decreases, as well as the height of the dense phase
    local maxima. In addition, the low-density phase becomes more populated and shifts towards  $\rho = 1.018$ 
    as $N_p/N$ increases. All of this suggests that the mixture is increasingly more 
    homogeneous when increasing the fraction of passive particles in the system, 
    a phenomenon also noticeable in the snapshots from Fig. \ref{fig:morfologias}.

Inspecting Fig. \ref{fig:mixphase} in more detail, it is interesting to note how the lever rule, which is frequently used to 
determine the fraction of high and low density phases at equilibrium in systems involving two phases, seems to work also in the 
case of mixtures of active and passive ABP. On the one hand, in Fig. \ref{fig:mixphase}(b), the point corresponding to 
$N_p/N=0.1$ has a density $\rho$ which is larger than the critical density $\rho_0$ corresponding to that fraction of passive 
particles, and thus is closer to the high density binodal curve. The accompanying $P(\rho)$ shows that a much higher fraction 
of particles can be found in a high density phase. On the other hand, the point corresponding to $N_p/N=0.7$ has a $\rho$ 
which is smaller than the corresponding critical density $\rho_0$. In that case, the related $P(\rho)$ shows that most 
particles can be found in a low density phase.


It has been already shown that in the case of a pure ABP suspension
the system is characterised by an effective diffusivity $D_e(\rho)$ that  depends on  density \cite{Fily_prl}.
 Thus, one should expect also an effective diffusivity for passive particles in a bath of active ones, which could depend not only on the self-propel velocity of the active particles but also on the total density  and  fraction of passive particles in the system.

\begin{figure}[h!]
\includegraphics[width=0.75\columnwidth]{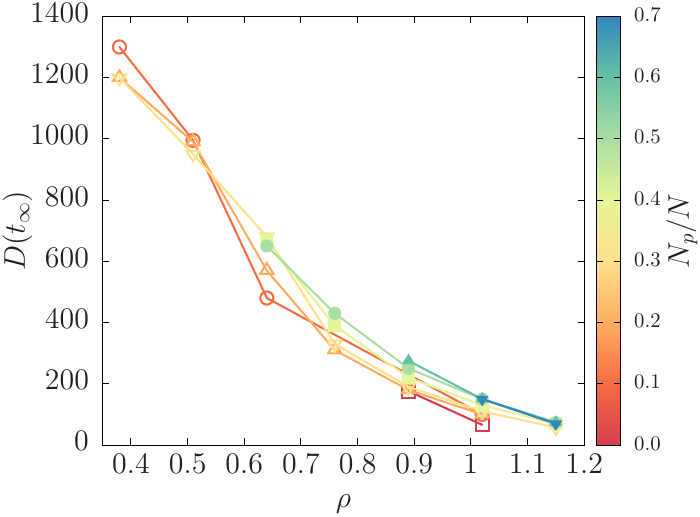}
\includegraphics[width=0.73\columnwidth]{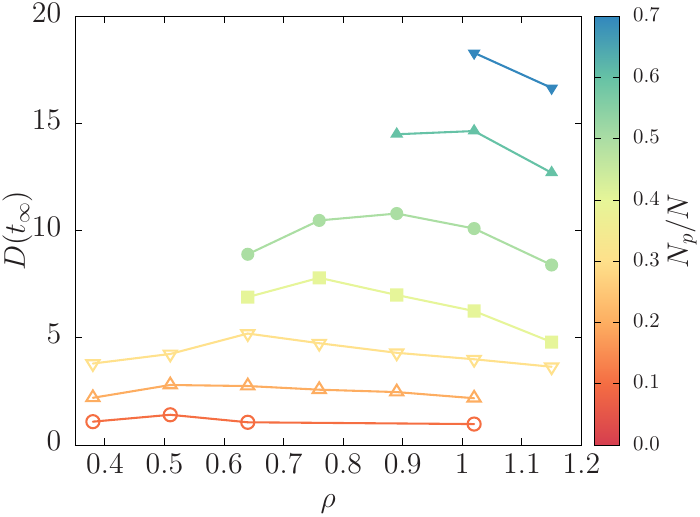}
\caption{Long time diffusion coefficient as a function of the total number density for active  (top) and passive particles (bottom) at Pe=120 and different passive particles concentration (as shown by the color code), corresponding to the state points shown in Fig,6 a).}
\label{fig:diff}
\end{figure}

In order to unravel how passive particles affect the dynamics of active ones (and viceversa) in the MIPS region, 
we compute the long-time diffusion coefficient from the MSD of each species at different 
passive particles concentrations (Fig.\ref{fig:diff}) 
fixing the value of the  Peclet  number ($Pe=120$). This value corresponds to the one already chosen in 
 Fig.6  a) to study  how the appearance of MIPS is  affected by the different $N_{p}/N$. 


The top panel of Fig.\ref{fig:diff} represents the active particles' diffusion  as a function of density.
At the density at which  MIPS appears, either in purely active (red) or in systems with $N_p/N$ up to about 0.5  (light green), 
the diffusion coefficient changes its slope and 
we detect different scenarios. 
At low density, 
where MIPS does not take place, active particles freely diffuse 
(the lower the density the higher  the diffusion). When MIPS starts appearing in the system
(as an example, $\rho \sim 0.5$ in the $N_{p}/N=0.1$ case), many active particles 
move to  the  boundaries of the densest phase and their diffusion coefficient decreases (the higher the density the slower particles diffuse). 
When the presence of MIPS is debatable ($N_{p}/N=0.7$ and high densities), active particles diffuse very slowly. 

The bottom panel of Fig.\ref{fig:diff} represents the passive particles' diffusion coefficient at the same conditions as in panel a, having set the same $Pe=120$ for active particles. 
To start with, the passive particles' diffusion is almost two order of magnitude lower than the active particles' one.
When the density increases, as soon as the system enters into the MIPS region, 
the diffusion coefficient shows a non-monotonic behaviour 
at the same density where 
the active particles diffusivity shows an inflection point. 
This effect is barely visible at very low $N_{p}/N$, due to the small number of passive 
particles in the system, but becomes clearer as soon as $N_{p}/N=0.3$. 
A possible explanation for this is the following.
At low densities the system 
is not phase separated  yet, and the diffusivity of passive particles 
increases as the density increases until the latter reaches the value corresponding to the MIPS boundary. 
When the system enters a MIPS region, while active particles start slowing down, 
passive particles' diffusivity increases as an effective result of the active particles' speed. 
Once in the MIPS phase, at higher densities, 
passive particles are trapped by a layer formed by the active ones: this is the reason why their diffusivity decrease.
Therefore, we conclude that either an inflection point in the active particles diffusivity (especially at low 
$N_{p}/N$) or a maximum of the passive particles diffusivity (especially at higher $N_{p}/N$), when plotted as a function of the total particle density $\rho$, could be used as a dynamical fingerprint for the appearance of MIPS in an active/passive binary mixture. 





When active particles are swimmers and passive particles are tracers five times smaller than the swimmers,
 in the dilute regime \textit{i.e.} ($\rho_{active} < 0.06$ and $\rho_{passive}=0.0034$), Delmotte et al. \citep{Delmotte}
  have found that the onset of the diffusive regime occurs after a period of anomalous transport at short times, 
  as the active packing fraction of swimmers increases. 
  Even though we are in a more concentrated regime without hydrodynamics and passive particles have the same size as the active particles, 
  we have also found a period of anomalous transport at short times as the number of passive particles increases (not shown here). 
  When studying the system with hydrodynamics,   the effective diffusion coefficient of tracers, calculated in the diffusive regime  at long times, increases with the swimmers packing fraction in the same way as in the experiments \citep{Delmotte}. 
   In contrast, without hydrodynamics and in more concentrated regimes, we have found the effective diffusion coefficient of tracers showing a non-monotonic behaviour due to MIPS, see bottom panel of Fig. \ref{fig:diff}

\begin{figure}[h!]
\includegraphics[width=\columnwidth]{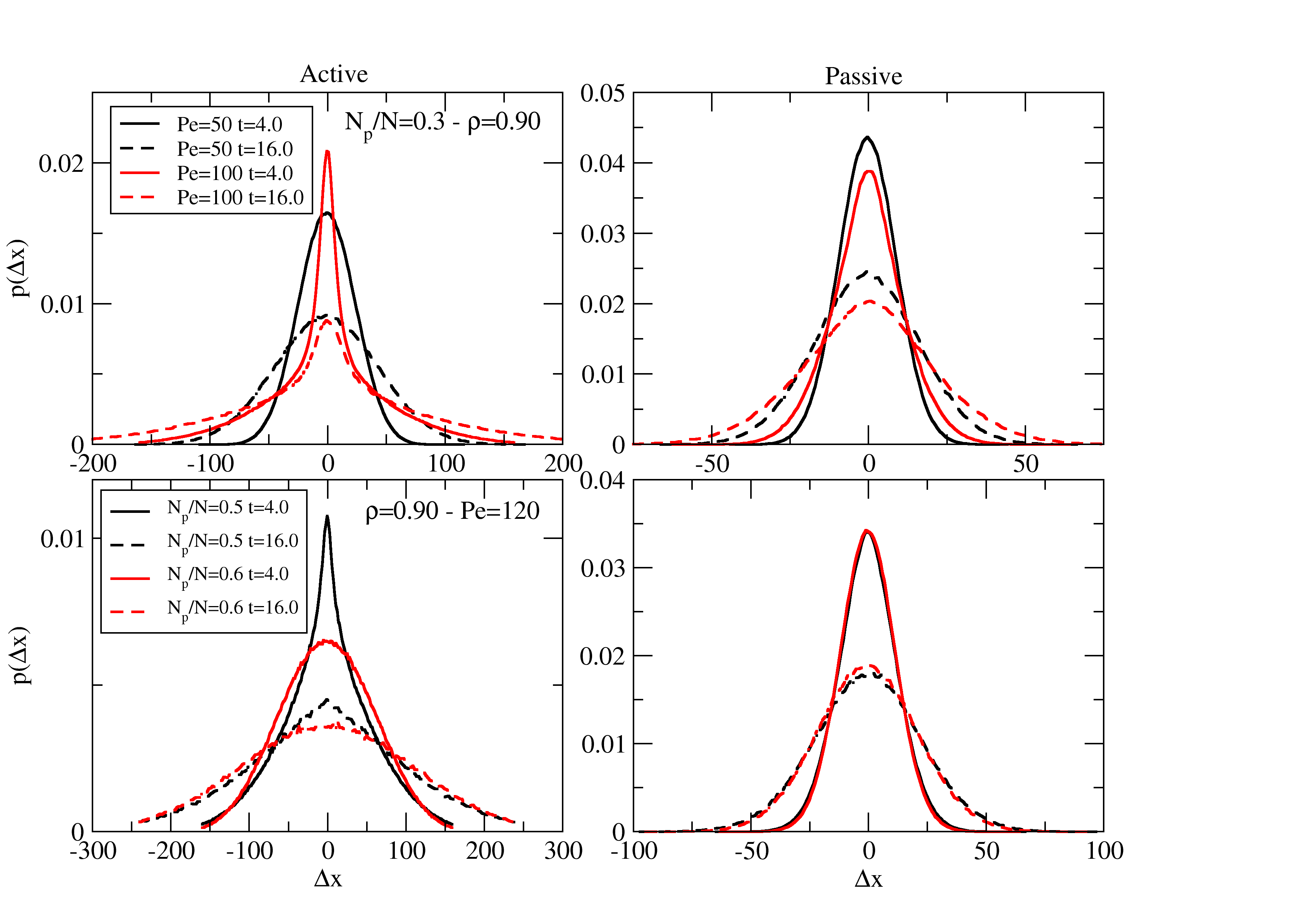}
\caption{
Probability distribution of the displacements along the $x$ axis measured at different times of active (left) and passive (right panels), in mixtures with $N_p/N=0.3$ and $\rho=0.9$ (top) for Pe=50 (no MIPS) and Pe=100 (MIPS), and in mixtures with $\rho=0.9$ and Pe=120 for $N_p/N=0.5$ (MIPS) and $N_p/N=0.6$ (no MIPS).}
\label{fig:pdf}
\end{figure}

To further study the dynamics of the mixture, we compute the distribution of particle displacements for both active and passive particles in the steady state (Fig. \ref{fig:pdf}).
   
In Fig. \ref{fig:pdf} top panels we focus on the system at  $N_{p}/N=0.3$ and density $\rho=0.90$ outside (Pe=50, in black) and inside (Pe=100, in red) the MIPS state, comparing 
the $P(\Delta x)$ at short (continuous) and long (dashed) times for both active (left-side) and passive (right-side) particles. 

In the passive case, independently on whether the system is  in the MIPS state, the $P(\Delta x)$  is always a Gaussian distribution centred in $\Delta x = 0$ whose width increases with time. The same happens for active particles when the system is outside a MIPS state. On the contrary, even though the time evolution is the same, the shape of the active particles' $P(\Delta x)$ is clearly non-Gaussian when the system is in a MIPS state. 
When computing the PDF at intermediate times, active particles in the less dense phase reminds the behaviour of particles  following Levy flight dynamics, 
showing the PDF a well-defined cusp and slowly decaying tails. When averaging with the contribution coming from active particles in the more dense phase, these characteristics are a  smeared out but still present. 

In order to underline the dependence of our results on the passive particles' concentration, we repeat the same 
study choosing a state point where the system is in a MIPS state ($N_{p}/N=0.5$, in black), compared to the same state point of a system which is right outside the MIPS state ($N_{p}/N=0.6$, in red),  compute 
the $P(\Delta x)$ at short (continuous) and long (dashed) times for both active (left-side) and passive (right-side) particles(bottom panels in Fig. \ref{fig:pdf}), and  compare to the same state point of a system which is right outside the MIPS state ($N_{p}/N=0.6$, in red). 
Once more, the $P(\Delta x)$  for passive particles is always a Gaussian, whose width increases with time.   
However, in the active case we detect clear differences, given that the distribution 
 is clearly not Gaussian 
 when the system is in a MIPS state (in black) and becomes Gaussian increasing the number of passive particles thus exiting a MIPS state.

Therefore, instead of detecting MIPS by computing the local density (via Voronoi tessellation), 
we suggest a   different route based on dynamics.  
When computing the $P(\Delta x)$ of active particles at intermediate times, when its shape changes from being Gaussian to being non-Gaussian, the system enters a MIPS state. At very long times, active particles move between different clusters, exploring the low density region along the way, and $P(\Delta x)$ returns to a Gaussian.

Delmotte et al. in \citep{Delmotte},  to study the effect of  steric versus hydrodynamic interactions, have carried out simulations of bidisperse suspensions without hydrodynamic interactions and with and without self-propulsion. The complete passive case developed a Gaussian PDF as expected, while the active/passive case without hydrodynamics developed a PDF with non-Gaussian tails. In our case, perhaps the non-Gaussian tails are not observed at this time-scales, but we did see wider tails in our PDFs. This effect  might be due to the tracers' size  and the simulated densities. Thus, in our case active particles have less entrainment of passive ones, than in the case of Ref. \citep{Delmotte} with smaller tracers.

As suggested in supercooled liquids\cite{dynhet}, a non-Gaussian behaviour of  $P(\Delta x)$ corresponds to 
dynamic heterogeneities in the system, i.e. spatially separated regions of mobile and less mobile particles. 
When MIPS occurs, most of active particles slow down forming the boundary of the phase separated 
region, whereas the remaining keeps propelling in the system.  
Therefore, a change in the shape of $P(\Delta x)$ is linked to a change in particles' dynamics.



\section{Conclusions}
In this work we have carried out an extensive study on the role of passive components in  
presence of ABP for a purely repulsive interacting potential (WCA) in two dimensions. 
Firstly, we have elaborated state diagrams for several fractions of passive components, 
 unveiling the necessary conditions for MIPS to occur in these binary mixtures.
 We have found that 
passive agents tend to hinder phase separation, but MIPS can still be found in binary mixtures, 
even for contents of passive colloids as high as $70 \%$. 
All MIPS regions can be scaled on top of another, culminating in what could be identified as a critical point.

Given that the rescaled state diagrams are similar, we studied the structures of MIPS. 
The underlying mechanism involves 
the self-assembly of active fronts that encapsulate passive bulks, giving rise to a phase separation 
for which dense clusters with a skin of active agents surrounding a mainly passive core are found. 
Thus we observe that active colloids can induce structure of passive components and trigger
  crystallisation. 
To better characterise the system's structure, we compute  the local number density distribution
function, whose shape is strongly affected by the amount of active particles present in the system.
We conclude that the mixture is increasingly more homogeneous  when increasing the fraction 
of passive particles. 

In order to study how passive particles affect the dynamics of the mixture, we compute the long-time 
diffusion coefficient. Interestingly, at the density at which the system enters a MIPS state the active particles' diffusivity 
shows an inflection point and the passive particles' diffusivity goes through a maximum, due to the change in the dynamics of the active components clearly described by the  $P(\Delta x)$. 



We believe that 
our results could help understanding the phenomena occurring in 
suspensions of particles with heterogeneous activity, shedding light upon the behaviour of active matter
 in complex media, and can be used to refine microrheological measurements.

\section*{Conflicts of interest}
There are no conflicts to declare. 

\section*{Acknowledgements}
The authors acknowledge funding from Grant
FIS2016-78847-P of the
MINECO and the UCM/Santander PR26/16-10B-2.
Francisco Alarc\'on acknowledges support from the ``Juan de la Cierva''
program (FJCI-2017-33580).
Raul Martinez acknowledge funding from MINECO (Ministerio de Econom\'ia, Industria y Competitividad, Spain) FPI grant BES-2017-081108.
 The authors acknowledge the
computer resources and technical assistance provided by the
Centro de Supercomputaci\'on y Visualizaci\'on de Madrid
(CeSViMa).






\section{Appendix}

\subsection{Different formation of MIPS for equal P\'eclet numbers \label{sec:peclet}}

We have found disagreement between the phase diagram given by Stenhammar \textit{et al.} \cite{Stenhammar} and our phase diagram for purely active systems with WCA interactions. We have observed that the discrepancy is due to our different approach for varying the P\'eclet number. In this paper $D_r$ is set constant and the self-propelling speed is changed to alter the P\'eclet number, while Stenhammar \textit{et al.} vary Pe by fixing $v$ and adjusting $D_r$. We have recovered the results from Stenhammar \textit{et al.} by following their methodology. From an energetic perspective, Pe measures the ratio between the ballistic energy $|F_a| \sigma$ and the thermal energy $k_B T$. Since we are employing a WCA potential, the effective diameter for the interactions $\sigma_{\text{eff}}$ is not the same for all collision energies due to the softness of the potential. If colloids were ideal hard spheres, $\sigma_{\text{eff}}$ would be the same for all self-propelling forces of active agents $|F_a|$. This means that, for WCA, changing Pe by varying $|F_a|$ effectively softens the particles as $|F_a|$ grows, reducing their effective diameter and therefore hindering self-assembly and interfering in MIPS, since phase separation would not be found in the limit of point ABP. This presumably explains the differences between both phase diagrams for ABP. 

\

Another aspect to consider in this regard is the role of thermal phenomena in the competition between diffusive behavior and self-propulsion effects, as presented in Section \ref{sec:intro}. Decreasing $D_r$ lessens the mean frecuency in which particle orientation is reset, implying an increase in the amount of time a trapped particle has to wait until it can have a chance to leave the rim of its cluster. In the limit of no rotational diffusion, a blocked active particle will remain pushing towards the cluster, and therefore contributing to its growth. Thus one would expect MIPS to be favored due to the cancellation of diffusive effects when increasing the P\'eclet number by decreasing $D_r$.

\

However, our method suits the one used in previous works, both experimental \cite{mips, experiment} and computational \cite{reviewpe, reviewpe2, analytical}, since it is more controllable and reproducible to change the self-propelling velocity (or, equivalently, $|F_a|$) in experiments by changing the chemical properties of the system in which active colloids are interacting than modifying the rotational diffusion. This suggests the need for an extensive study on the characterization of the P\'eclet number for out of equilibrium systems and for soft potentials, since this number is widely used to describe the competetition between active and diffusive motion \cite{bechinger, mix1, mix2, Stenhammar} but it seems that Pe is not sufficient to describe these systems, since potential steepness comes into play.

\end{document}